\documentclass[12pt]{article}
\begin{document}
\baselineskip=7mm
%%%%%%%%%%%%%%%%%%%%%%%%%%%%%%%%%%%%%%%%
%%    Macros
%%%%%%%%%%%%%%%%%%%%%%%%%%%%%%%%%%%%%%%%
\renewcommand{\baselinestretch}{1.28}
\newcommand{\ba}{\begin{eqnarray}}
\newcommand{\ea}{\end{eqnarray}}
\newcommand{\non}{\nonumber\\}

%%%     only for thebibliography
\makeatletter
\def\section{\@startsection{section}{1}{\z@}
    {-3.5ex plus -1ex minus -.2ex}{2.3ex plus .2ex}
    {\hskip 63.1mm \large\bf}}
\makeatother

%%%%%%%%%%%%%%%%%%%%%%%%%%%%%%%%%%%%%%%%
%%    Titlepage
%%%%%%%%%%%%%%%%%%%%%%%%%%%%%%%%%%%%%%%%
\thispagestyle{empty}
\hskip 10cm{}
\vspace{2cm}

\centerline{\LARGE  Information metric on instanton moduli spaces}
\vskip 2mm
\centerline{\LARGE in nonlinear $\sigma$ models}

\vspace{1.5cm}

\centerline{\bf Shigeaki Yahikozawa \footnote{
e-mail: {\tt yahiko@rikkyo.ac.jp}}}
\vspace{1mm}

\centerline{\it Department of Physics, Rikkyo University}
\centerline{\it Tokyo 171-8501, Japan}
\vspace{1cm}

\centerline{\bf Abstract}
\vspace{2mm}
We study the information metric on instanton moduli spaces in 
two-dimensional nonlinear $\sigma$ models.
In the CP${}^1$ model, the information metric on the moduli space
of one instanton with the topological charge $Q=k\,\, (k\geq 1)$ is a 
three-dimensional hyperbolic metric, which corresponds to Euclidean anti--de 
Sitter space-time metric in three dimensions, and the overall scale factor 
of the information metric is $4k^2/3$; this means that the sectional
curvature is $-3/4k^2$.
We also calculate the information metric in the CP${}^2$ model.

%%%%%%%%%%%%%%%%%%%%%%%%%%%%%%%%%%%%%%%%%
%%    Introduction
%%%%%%%%%%%%%%%%%%%%%%%%%%%%%%%%%%%%%%%%%
\newpage
\setcounter{footnote}{0}
\pagestyle{plain}
\pagenumbering{arabic}
\noindent
{\large \bf 1. Introduction}
\vspace*{2mm}

A parametrized family of probability distributions is often treated as a
statistical model, which is deeply related to nonlinear $\sigma$ (NL$\sigma$)
models. The NL$\sigma$ models have arisen 
in various contents \cite{Jacobson}--\cite{Cardy}: 
the Heisenberg ferromagnet model, the quantum Hall 
effect, and other statistical mechanics problems; and the conformal field theory,
string theory, and  Yang-Mills (YM) theories in four dimensions. 
In such models, instantons play an important role in nonperturbative analyses 
\cite{Kamenev}--\cite{Patrascioiu}.
{}From the geometrical point of view, the information metric arises as a metric
on the moduli space of the instantons, and in more general as a metric on 
a manifold of probability distributions. It is generally defined by
\ba
\label{inf-met}
G_{AB}=\int d^D\!x \,\,p(x;\theta) 
                 \,\partial_A\!\ln p(x;\theta) \,\,\partial_B\!\ln p(x;\theta) \,\,\,,
\ea
where $p(x;\theta)$ is the probability density function of $x$,
parametrized by $\theta$ \cite{Amari-0}\cite{Amari}.
Here,  $x$ is assumed to belong to a flat $D$-dimensional
space ${\mbox{\boldmath ${\bf R}$}}^D$, 
$\theta$ is a real $N$-dimensional parameter
$(\theta_1, \theta_2, ... , \theta_N)$, and $\partial_A$ 
is the derivative with respect to the parameters
$\theta_A\, (A=1, 2, ... , N)$.  The space of the parameters corresponds
to the instanton moduli space in this paper.
If we treat the normalized topological charge density of the instantons, which 
is the same as the normalized energy density, 
as the probability density function, the information metric on the 
moduli space of the instantons naturally appears from the definition
 (\ref{inf-met}).

Geometrical approaches have many advantages in statistical physics
\cite{Ruppeiner-0}--\cite{Ruppeiner-1}, for example, 
it has recently been shown that the scalar curvature of the information metric
plays a central role in studying  the characterization of the phase structure
in statistical mechanics models \cite{Janke-0}--\cite{Janyszek}. 
In this paper, we concentrate upon 
the geometrical perspective from which instantons in nonlinear $\sigma$
models are combined with the information metric.
The information metric on the moduli space of the instantons has been
studied in YM theories in four dimensions and in
NL$\sigma$ models in two dimensions. It has been shown so far
that the information metric is isometric to hyperbolic space
when one instanton with topological charge 1 exists in a $SU(2)$ YM theory
\cite{Hitchin}--\cite{Blau}
or in a rational map \cite{Murray} which is correspondent to 
a NL$\sigma$ model in two dimensions, and also that the information metric is 
nondegenerate for the moduli space of the multiinstantons in rational maps
\cite{Murray}. 
What has not been done so far is to study the concrete dependence of the
information metric on  the topological charge of the instantons. 
That is what we shall do in this paper. 
We shall, in particular, compute the overall scale factor, which corresponds to 
the square of ``radius," of the information metric in the moduli space of one 
instanton with the topological charge which is greater than or equal to 1.
The negative reciprocal of the overall scale factor of the information metric 
is directly correspondent to  the curvature, which is the  sectional curvature 
to be exact. 
In the geometrical analyses about the NL$\sigma$ models and statistical models 
in general, it is important to investigate the curvature of the moduli spaces
where the instantons live.
 
The paper is organized as follows. In Sec. 2, we briefly review 
the CP${}^n$ NL$\sigma$ model. In Sec. 3, we compute the information 
metric on the moduli space of one instanton with the topological charge which 
is greater than or equal to 1 in the CP${}^1$ model. And, furthermore,
we compute the information metric in the CP${}^2$ model.
In Sec. 4, we draw conclusions and indicate some interesting possibilities 
to extend our work.

\vspace{8mm}
%%%%%%%%%%%%%%%%%%%%%%%%%%%%%%%%%%%%%%%%%
%%    Definition of the CP^n nonlinear sigma model
%%%%%%%%%%%%%%%%%%%%%%%%%%%%%%%%%%%%%%%%%
\noindent
{\large \bf 2.  Definition of the CP${}^n$ NL$\sigma$ model}
\vspace*{2mm}

To define the two-dimensional CP${}^n$ model, we take an $(n+1)$-dimensional 
complex vector field $\Phi$ \cite{D'Adda}\cite{Furuta}:
\ba
\Phi=(\phi_1, \phi_2, ... , \phi_{n+1}) \,\,\,.
\ea
The CP${}^n$ model is defined by the Lagrangian density $\cal L$
 in two-dimensional Euclidean space-time:
\ba
\label{Lagrangian}
{\cal L}=\frac{1}{2g^2} [(D_\mu\Phi)(D_\mu\Phi)^{\dag}
+\alpha(\Phi\Phi^{\dag}-1) ] \,\,\,,
\ea
where $D_\mu$ $(\mu=1, 2)$ is the covariant derivative defined by
\ba
D_\mu&=&\partial_\mu+iA_\mu \,\,\,, \\
A_\mu&=&\frac{i}{2} [\Phi(\partial_\mu\Phi)^{\dag}
-(\partial_\mu\Phi)\Phi^{\dag} ] \,\,\,,
\ea
$\alpha$ is the multiplier field imposing the constraint
$\Phi\Phi^{\dag}=1$, and $g$ is the coupling constant.
This model has the global $SU(n+1)$ symmetry and the $U(1)$ local 
symmetry.

The CP${}^n$ model defined above has a topological charge $Q$, which is also
called a winding number:
\ba
Q&=&\int d^2x \, q \,\,\,, \\
q&=&-\frac{1}{2\pi}\epsilon_{\mu\nu}\partial_{\mu}A_{\nu}
\non
&=&-\frac{i}{2\pi}\epsilon_{\mu\nu}
(D_{\mu}\Phi)(D_{\nu}\Phi)^{\dag} \,\,\,,
\ea
where $q$ is the topological charge density.

The equation of motion is obtained from the Lagrangian density (\ref{Lagrangian}),
\ba
D_{\mu}D_{\mu}\Phi+(D_{\mu}\Phi)(D_{\mu}\Phi)^{\dag}\Phi=0\,\,\,.
\ea
If the self-dual equation
\ba
\label{self-dual}
D_{\mu}\Phi=-i\epsilon_{\mu\nu}D_{\nu}\Phi
\ea
is satisfied, then the equation of motion is automatically satisfied.
Using this self-dual equation, we get the topological charge density $q$ as follows:
\ba
q=\frac{1}{2\pi}(D_{\mu}\Phi)(D_{\mu}\Phi)^{\dag} \,\,\,,
\ea
which is proportional to the Lagrangian density under the constraint
$\Phi^{\dag}\Phi=1$, namely, the relation between the Lagrangian density 
and the topological charge density is
\ba
{\cal L}=\frac{\pi}{g^2} q \,\,\, .
\ea

\vspace{8mm}
%%%%%%%%%%%%%%%%%%%%%%%%%%%%%%%%%%%%%%%%%
%%    Information metric on the instanton moduli space
%%%%%%%%%%%%%%%%%%%%%%%%%%%%%%%%%%%%%%%%%
\noindent
{\large \bf 3. Information metrics on moduli spaces of instantons}
\vspace*{2mm}

To find the self-dual solution, let us parametrize the field $\Phi$ as follows
\ba
\Phi&=&\frac{W}{\sqrt{WW^{\dag}}} \,\,\,, 
\ea
where $W$ is an $(n+1)$-dimensional vector,
\ba
W&=&(w_1, w_2, ... , w_{n+1}) \,\,\, .
\ea
Substituting this $\Phi$ into the self-dual equation (\ref{self-dual}), we find that
if $\partial_{\bar{z}}W=0$, then the self-dual equation is satisfied, where
$z=x^1+ix^2$. Moreover, by making use of the $U(1)$ local symmetry, we can take
$W$ as $W=(\eta, u_2, u_3, ..., u_{n+1})$, where $\eta$ is a real
number $(\eta\geq0)$ and $u_i$'s$ $\,$(i=2, 3, ..., n+1)$ are rational functions.

\vspace{5mm}
%%%%%%%%%%%%%%%%%%%%%%%%%%%%%%%%%%%%%%%%%
%%    CP^1 model
%%%%%%%%%%%%%%%%%%%%%%%%%%%%%%%%%%%%%%%%%
\noindent
{\large 3-1. Information metric in the CP${}^1$ model}
\vspace*{2mm}

Let us consider the instanton solution in the CP${}^1$ model, which is the most 
simple NL$\sigma$ model.
Since the purpose of this paper is to investigate the information metric 
on the moduli space of the instanton with the topological charge which is greater 
than or equal to 1, we adopt the following self-dual solution:
\ba
\label{inst-sol}
W=[{\lambda}^k, (z-a)^k]\,\,\,,
\ea
where $a=a^1+ia^2$ and $k$ is any positive integer. 
Here, $\lambda$ and $a$ correspond to instanton moduli parameters. 
The topological charge density of this instanton solution is given by
\ba
q=\frac{k^2}{\pi} 
     \frac{{\lambda}^{2k}|z-a|^{2k-2}}{({\lambda}^{2k}+|z-a|^{2k})^2} \,\,\, ,
\ea
which has a maximum value where $|z-a|=[(k-1)/(k+1)]^{1/2k}\lambda$\,
$(k\geq1)$. The topological charge of the instanton solution (\ref{inst-sol}) is $k$:
$Q=k$. 
We take the normalized topological charge density,
which corresponds to the normalized energy density or the normalized Lagrangian 
density in the Euclidean space-time, as the probability density function
$p(z; \lambda, a)$ of the instanton in the CP${}^1$ model:
\ba
\label{probability}
p(z; \lambda, a)&=&\frac{1}{ k} q
\non
&=&\frac{k}{\pi} 
\frac{{\lambda}^{2k}|z-a|^{2k-2}}{({\lambda}^{2k}+|z-a|^{2k})^2} \,\,\,.
\ea
By substituting this probability density function into the definition of  
the information metric $(\ref{inf-met})$, we obtain 
\ba
ds^2&=&G_{AB}d{\theta}^Ad{\theta}^B\,\,\,,
\ea
where
\ba
G_{11}&=&\frac{4}{3}  k^2  \frac{1}{{\lambda}^2} \,\,\,, \\
G_{ij}&=&\frac{2\pi}{3}  \frac{k^2-1}{k\sin(\pi/k)} 
\frac{1}{{\lambda}^2}  \delta_{ij} \,\,\,, \\
G_{1i}&=& G_{i1}=0 \,\,\,,
\ea
where ${\theta}^1=\lambda, {\theta}^2=a^1, {\theta}^3=a^2$
and $i, j=2, 3$.
If we change the instanton moduli parameter $a$ into $b$ as follows
\ba
a=\sqrt{\frac{2}{\pi} \frac{k^3\sin(\pi/k)}{k^2-1}}   b \,\,\,,
\ea
where $b=b^1+ib^2$,
it is established that the information metric on the moduli space 
$(\lambda, \vec{b})$ of the instanton is
\ba
ds^2=\frac{4k^2}{3} \frac{d{\lambda}^2 + d{\vec{b}}^2}
{{\lambda}^2} \,\,\,,
\ea
where $\vec{b}=(b^1, b^2)$.  The overall scale factor of the information metric
is $4k^2/3$. This hyperbolic three-space corresponds 
to the Euclidean anti--de Sitter space-time in three dimensions with 
the ``radius" $R=\sqrt{4/3}\,k$, which means that the sectional curvature is 
$-3/4k^2$. If $k$ becomes very large, the spacetime becomes flatter. 
It is clear that the information metric on the moduli space of 
one anti-instanton with the topological charge which is any negative integer is 
also the three-dimensional hyperbolic metric with the same curvature.

\vspace{5mm}
%%%%%%%%%%%%%%%%%%%%%%%%%%%%%%%%%%%%%%%%%
%%    CP^2 model
%%%%%%%%%%%%%%%%%%%%%%%%%%%%%%%%%%%%%%%%%
\noindent
{\large 3-2. Information metric in the CP${}^2$ model}
\vspace*{2mm}

In the CP${}^2$ model, we consider the information metric on the moduli
space of instantons which have $Q=1$ and moduli parameters ($\lambda$,
$a$, $b$). The instanton solution is
\begin{equation}
W=(\lambda, z-a, z-b)\,\,\,,
\end{equation}
where $\lambda$ is real, $a$ and $b$ are complex numbers. In this case,
the probability density function $p(z; \lambda, a, b)$ is given by
\begin{equation}
p(z; \lambda, a, b)=\frac{4}{\pi} \frac{2\lambda^2+|a-b|^2}
{(2\lambda^2+|a-b|^2+|2z-a-b|^2)^2}\,\,\,.
\end{equation}
If we change the coordinate $z$ and the moduli parameters ($\lambda$,
$a$, $b$) into $v$ and ($\Lambda, c, \delta$) as follows
\begin{eqnarray}
\label{change-1}
z&=&\frac{1}{\sqrt{2}}v\,\,\,, \\
\label{change-2}
a&=&\frac{1}{\sqrt{2}}(c+\delta)\,\,\,, \\
\label{change-3}
b&=&\frac{1}{\sqrt{2}}(c-\delta)\,\,\,, \\
\label{change-4}
\lambda&=&\sqrt{\Lambda^2-\delta^2}\,\,\,,
\end{eqnarray}
the probability density
function becomes
\begin{equation}
\tilde{p}(v; \Lambda, c, \delta)=\frac{1}{\pi} 
\frac{\Lambda^2}{(\Lambda^2+|v-c|^2)^2}\,\,\,.
\end{equation}
This is indeed the same as the probability density function
for one instanton with $Q=1$ in the CP${}^1$ model [see Eq. (\ref{probability})].
Therefore, the information metric of the moduli space ($\Lambda, c, \delta$)
is easily obtained,
\begin{equation}
\label{metric2}
ds^2=\frac{4}{3} \frac{d\Lambda^2+d\vec{c}^2}{\Lambda^2}\,\,\,,
\end{equation}
where $\vec{c}=(c^1, c^2)$. 
This information metric represents the hyperbolic three-space, which 
corresponds to the Euclidean anti--de Sitter space-time in three dimensions, 
in the same way as the case of the CP${}^1$ model mentioned in Sec. 3-1.
Notice that there is no dependence on the parameters $\delta$ and 
$d\vec{\delta}$ in this information metric, 
where $\vec{\delta}=(\delta^1, \delta^2)$.
The substitution of Eqs. (\ref{change-1})--(\ref{change-4}) into 
Eq. (\ref{metric2}) leads to the information metric
for the moduli parameters ($\lambda, a, b$):
\begin{eqnarray}
ds^2&=&G_{AB}d\theta^A\theta^B\,\,\,, \\
G_{11}&=&\frac{16}{3} \frac{\lambda^2}{(2\lambda^2+|a-b|^2)^2}\,\,\,, \\
G_{1i}&=&G_{i1}=-G_{1\, i+2}=-G_{i+2\, 1} \non
&=&\frac{8}{3} \frac{\lambda\tilde{\delta}_i}
{(2\lambda^2+|a-b|^2)^2}\,\,\,, \\
G_{ij}&=&G_{ji}=G_{i+2\, j+2}=G_{j+2\, i+2} \non
&=&\frac{4}{3} \frac{\tilde{\delta}_i\tilde{\delta}_j}
{(2\lambda^2+|a-b|^2)^2}+\frac{4}{3} \frac{\delta_{i j}}
{2\lambda^2+|a-b|^2}\,\,\,, \\
G_{i\, j+2}&=&G_{j+2\, i} \non
&=&-\frac{4}{3} \frac{\tilde{\delta}_i\tilde{\delta}_j}
{(2\lambda^2+|a-b|^2)^2}+\frac{4}{3} \frac{\delta_{i j}}
{2\lambda^2+|a-b|^2}
\end{eqnarray}
where $\theta^1=\lambda, \,\theta^2=a^1,\, \theta^3=a^2,\, \theta^4=b^1, \,
\theta^5=b^2$,\,\, $i,j=2,3$,\,\, $A,B=1,2,\ldots,5$,\,\, 
and $\tilde{\delta}_i=a^i-b^i$.
Although this information metric looks complicated, a simple structure
is hidden as explained above.

\vspace{8mm}
%%%%%%%%%%%%%%%%%%%%%%%%%%%%%%%%%%%%%%%%%
%%    Information metric on the instanton moduli space
%%%%%%%%%%%%%%%%%%%%%%%%%%%%%%%%%%%%%%%%%
\noindent
{\large \bf 4. Conclusions and Discussions}
\vspace*{2mm}

We have shown that in the CP${}^1$ model the information metric on the moduli 
space of one instanton with the topological charge $Q=k\,\, (k\geq 1)$ is the 
Euclidean anti--de Sitter space-time metric in three dimensions.
The overall scale factor of the information metric is $4k^2/3$, and this means
that the ``radius" $R=\sqrt{4/3}\,k$ and the sectional curvature is 
$-3/4k^2$. If $k$ becomes very large, the space-time becomes flatter. 
Furthermore, we have also computed the information metric of the 
moduli space of the instanton in the CP${}^2$ model.

The NL$\sigma$ model often arises in string theory as well as in various statistical
mechanics problems.
Since the topological charge of the instanton may be related to  the Ramond-Ramond 
charge or the $D$-instanton charge in string theory, it is worthwhile to attempt to 
apply the above results to the correspondence between conformal field theories 
and string theories in the anti--de Sitter space-time, which is called  the AdS/CFT
 correspondence \cite{Maldacena}. 
It is particularly interesting to investigate the relation between 
three-dimensional anti--de Sitter space-time and a kind of two-dimensional 
NL$\sigma$ model from the point of view of string theory.
Furthermore, we should study the information metric on the moduli space of 
multiinstantons in the NL$\sigma$ models, the YM theories,  and supergravity 
theory in more detail.

\vspace{8mm}
%%%%%%%%%%%%%%%%%%%%%%%%%%%%%%%%%%%%%%%%%%
%%    Acknowledgment
%%%%%%%%%%%%%%%%%%%%%%%%%%%%%%%%%%%%%%%%%%
\noindent
{\large \bf Acknowledgments}
\vspace{3mm}

The present work was supported in part by Grant for the Promotion of Research and
Special Fund for Research from Rikkyo University, and Grant-in-Aid for Scientific 
Research (Grant No. 13127104) from the Japan Ministry of Education, Culture, Sports, 
Science and Technology.

%%%%%%%%%%%%%%%%%%%%%%%%%%%%%%%%%%%%%%%%%%
%     References
%%%%%%%%%%%%%%%%%%%%%%%%%%%%%%%%%%%%%%%%%%

\end{document}